\documentclass[5p]{elsarticle}
\usepackage{amsfonts}
\usepackage{setspace}
\usepackage{fixltx2e}
\usepackage{color}
\usepackage{subcaption}
\usepackage{epsfig}
\usepackage{url}
\usepackage{algorithm}
\usepackage{algorithmic}
\usepackage{color}
\usepackage{multirow}
\usepackage{mathtools}
\usepackage{lipsum}
\usepackage{dblfloatfix}
\usepackage{epstopdf}
\newcommand{\rank}{{\sf rank}}

\newcommand{\ds}{\displaystyle}

\newcommand{\la}{\langle}
\newcommand{\ra}{\rangle}

\newcommand{\clT}{{\mathcal T}}

\newcommand{\clL}{{\mathcal L}}
\newcommand{\clN}{{\mathcal N}}

\newcommand{\clG}{{\mathcal G}}

\newcommand{\clI}{{\mathcal I}}

\newcommand{\Tr}{{\sf Trace}}

\usepackage{lineno,hyperref}
\modulolinenumbers[5]

\journal{Journal of System $\&$ Control Letters}

\begin{document}

%
%
%
%
%
%
%
%

\begin{frontmatter}

\title{Global Optimal Power Flow over Large-Scale Power Transmission Networks\tnoteref{footnoteinfo}}

\tnotetext[footnoteinfo]{Corresponding author H.D. Tuan. Tel. +61.02.9405 9435. Fax +61.02.9405 9039}

\author[UTS]{Y. Shi}\ead{ye.shi@student.uts.edu.au}
\author[UTS]{H. D. Tuan}\ead{tuan.hoang@uts.edu.au}
\author[ONERA]{P. Apkarian}\ead{Pierre.Apkarian@onera.fr}
\author[UNSW]{A. V. Savkin}\ead{a.savkin@unsw.edu.au}

\address[UTS]{School of Electrical and Data Engineering,
University of Technology, Sydney, NSW 2007, Australia}
\address[ONERA]{CERT-ONERA, Control System Department, 2, avenua Edouard
Belin, 31055 Toulouse, France}
\address[UNSW]{School of Electrical Engineering and Telecommunications, the University  of New South
Wales, NSW 2052, Australia}

\begin{abstract}
Optimal power flow (OPF) over power transmission networks poses challenging  large-scale nonlinear optimization problems, which
involve a large number of quadratic equality and indefinite quadratic inequality constraints.
These computationally intractable constraints are often expressed by linear constraints plus
matrix additional rank-one constraints on the outer products of the voltage vectors.  The existing convex relaxation technique, which drops the difficult  rank-one constraints for tractable computation, cannot yield even a feasible point.
We address these computationally difficult problems by an iterative procedure,
which  generates a sequence of improved points that converge to a rank-one
solution. Each iteration calls a semi-definite program. Intensive simulations for the OPF
 problems over networks with a few thousands  of buses are provided to demonstrate the efficiency of our approach.
The suboptimal values of the OPF problems found by our
computational procedure turn out to be the global optimal value with computational tolerance less than 0.01\%.
\end{abstract}


\end{frontmatter}


\section{Introduction}
Smart grids are operated by the advanced distribution management system (DMS), which is responsible
for supervisory control and data acquisition in reactive dispatch, voltage
regulation, contingency analysis, capability maximization and other smart operations.
The optimal power flow (OPF) problem, which determines  a steady state operating point
that minimizes the cost of electric power generation or the transmission loss
is the backbone of DMS (see e.g. \cite{C62,HG91,MEA99,PJ08} and references therein).
Mathematically, the OPF problem is highly nonlinear and nonconvex due to numerous
quadratic equality and indefinite quadratic inequality constraints
for bus interconnections, hardware
operating capacity  and the balance between power demand and supply. These
nonlinear constraints are mathematically troublesome
so  the state-of-the-art nonlinear optimization solvers may  converge to just
stationary points (see \cite{Betal13} and references therein), which are not necessarily feasible.
To handle these nonlinear constraints, it is common
to reformulate them as linear constraints on the outer product $W=VV^H$ of the voltage vector
$V=(V_1, V_2,...,V_n)^T\in\mathbb{C}^n$. As a result, the OPF problem is
recast by a semi-definite program (SDP) plus the  additional rank-one constraint on
outer product matrix $W$ \cite{Betal08,LL12}. The rank-one constraint
is then dropped for  semi-definite relaxation (SDR). However, the optimal solution of SDR is
of rank-more-than-one in general and cannot help retrieval of a feasible point or stationary point  of the  OPF problem \cite{LL12,MSL15,MAL15}.
In \cite{STST15}, we have extended the technique of \cite{PTKN12} for solving the beamforming optimization problems in signal processing to optimize the  outer product matrix $W$, which works very well and is practical
for moderate-scale power distribution networks up to $n=300$ buses. There is another approach (see e.g.
\cite{J1} and references therein), which is based on hierarchies of moment-based relaxation for
nonconvex quadratic problems  to tackle large networks with simple nonconvex constraints.\\
Power transmission networks in modern smart grids are often devised  with a few thousand buses \cite{BYB06,HBB09,HHM11}.
Under a such large number $n$ of buses it is impossible to use the single matrix $W\in\mathbb{C}^{n\times n}$, which
involves $n(n+1)/2\approx O(10^{7})$ complex variables.
On the other hand, the number of the flow lines for bus connection is
relatively moderate  so only a small portion of the crossed nonlinear terms $V_kV_m^*$
appears in the nonlinear constraints. The common approach is to use the  outer products of
overlapped groups of the voltage variables to cover them \cite{Metal13,AHV14,MAL15}.
All rank-one constraints on these outer products are then dropped for SDR.
Obviously, the optimal solution of this SDR usually is not of rank-one
and thus does not have any physical meaning. There is no  technique to retrieve
a feasible rank-one point from the rank-more-than-one solution of SDR.\footnote{There is an algorithm
of finding a rank-one solution \cite[Alg. 1]{J2}, which however is applicable to simple nonconvex constraints
and is not guaranteed to convergence}
Multiple matrix rank constrained optimization
has received a great attention due to its potential application in robust control synthesis \cite{AT99a,ANP08}
but  to our best knowledge there is no effective computation so far. The contribution of this paper is
two-fold:
\begin{itemize}
\item An effective decomposition for large-scale OPF problems,
which involves essentially reduced numbers of the rank-one constraints on matrices of moderate size for expressing the network nonlinear constraints;
\item A new iterative procedure for  rank-one constrained optimization, which is
practical for computational  solutions of large-scale indefinite quadratic programming.
Simulations for the large-scale OPF problems show that it is capable of finding the global optimal
solution with the computational tolerance less than $0.01\%$.
\end{itemize}
The paper is structured as follows. Section 2 is devoted to the OPF problem formulation and its difficulties.
Its computational solution is developed in Section 3. Section 4 provides simulation to show the efficiency of our method. The conclusions are drawn in Section 5.

{\it Notation.} $j$ denotes the imaginary unit; $M\succeq 0$ means {\color{black}that}
 $M$ is a Hermitian symmetric positive semi-definite matrix; $\rank(M)$ is the rank of {\color{black}the} matrix $M$; $\Re (\cdot)$ and $\Im (\cdot)$ denote the real and imaginary parts of a complex quantity;
 $a\leq b$ for two complex numbers $a$ and $b$ is componentwise understood, i.e. $\Re(a)\leq \Re(b)$ and $\Im(a)\leq \Im(b)$;
$\la .,.\ra$ is the dot product of matrices, while ${\sf diag}\{A_i\}$ denotes the matrix with diagonal blocks $A_i$ and zero off-diagonal blocks; the cardinality of a set $\clL$ is denoted by $|\clL|$.
\section{Optimal power flow problem and challenges}
Consider an AC electricity transmission network with a set of $n$ buses ${\mathcal N}:=\{1,2,\cdots,n\}$.
The buses are connected through a set of flow lines ${\mathcal L}\subseteq {\mathcal N}\times {\mathcal N}$, i.e.
bus $m$ is connected to bus $k$ if and only if $(m,k)\in {\mathcal L}$. Accordingly,
$\clN(k):=\{m\in {\mathcal N}\ :\ (m,k)\in{\mathcal L}\}$. The power
demanded at bus $k\in\clN$  is $S_{L_k}=P_{L_k}+jQ_{L_k}$, where $P_{L_k}$ and $Q_{L_k}$ are the real and reactive power.
A subset ${\mathcal G}\subseteq {\mathcal N}$ of buses is supposed to be connected to generators.
Any bus $k\in {\mathcal N}\setminus {\mathcal G}$ is thus not connected to generators.

Other physical parameters are following \cite{Hbook,DT68,ZMT11}:
\begin{itemize}
\item $Y=[y_{km}]_{(k,m)\in \clN\times\clN}\in\mathbb{C}^{n\times n}$ is the admittance matrix \cite{ZMT11}. Each $y_{km}$ is
the mutual admittance between bus $k$ and bus $m$, so $y_{km}=y_{mk}$ $\forall\ (k,m)\in\clL$.
\item $V$ is the complex voltage vector, $V = [V_1,V_2,\cdots,V_n]^T\\\in \mathbb{C}^n$, where $V_k$ is the complex voltage injected to bus $k\in {\mathcal N}$.
\item $I$ is the complex current vector, $I = YV = [I_1,I_2,\cdots\\I_n]^T\in \mathbb{C}^n$, where $I_k$ is the complex current injected to bus $k\in {\mathcal N}$.
\item $I_{km}$ is the complex current in the power line $(k,m)\in {\mathcal L}$, $\ds \sum_{m\in {\mathcal N}(k)}I_{km} = I_k = \ds\sum_{m\in {\mathcal N}(k)} y_{km} V_m$.
\item $S_{km}=P_{km}+jQ_{km}$ is the complex power transferred from bus $k$ to bus $m$, where $P_{km}$ and $Q_{km}$ represent the
 real and reactive transferred power.
\item $S_{G_k}=P_{G_k}+jQ_{G_k}$ is the complex power injected by bus $k\in {\mathcal G}$, where $P_{G_k}$ and $Q_{G_k}$ represent the
 real and reactive generated power.
\end{itemize}

For each bus $k$, it is obvious that
\[
\begin{array}{lll}
S_{G_k} - S_{L_k}&=&(P_{G_k} - P_{L_k})+ j(Q_{G_k} - Q_{L_k})\\
&=&V_k I_k^*=V_k \ds \sum_{m\in {\mathcal N}(k)} V_m^*y_{km}^*.
\end{array}
\]
Therefore, the real generated power $P_{G_k}$ and reactive generated power $Q_{G_k}$ at bus $k$ are the following
nonconvex quadratic functions of the bus voltage vector variable $V:=(V_1, V_2,...,V_n)^T\in \mathbb{C}^n$:
$P_{G_k} = P_{L_k} + \Re (\ds \sum_{m\in {\mathcal N}(k)}V_k V_m^*y_{km}^*)$ and
$Q_{G_k} = Q_{L_k} + \Im (\ds \sum_{m\in {\mathcal N}(k)}V_k V_m^*y_{km}^*)$.

\begin{figure}[h]
\centering
\includegraphics[width=\columnwidth]{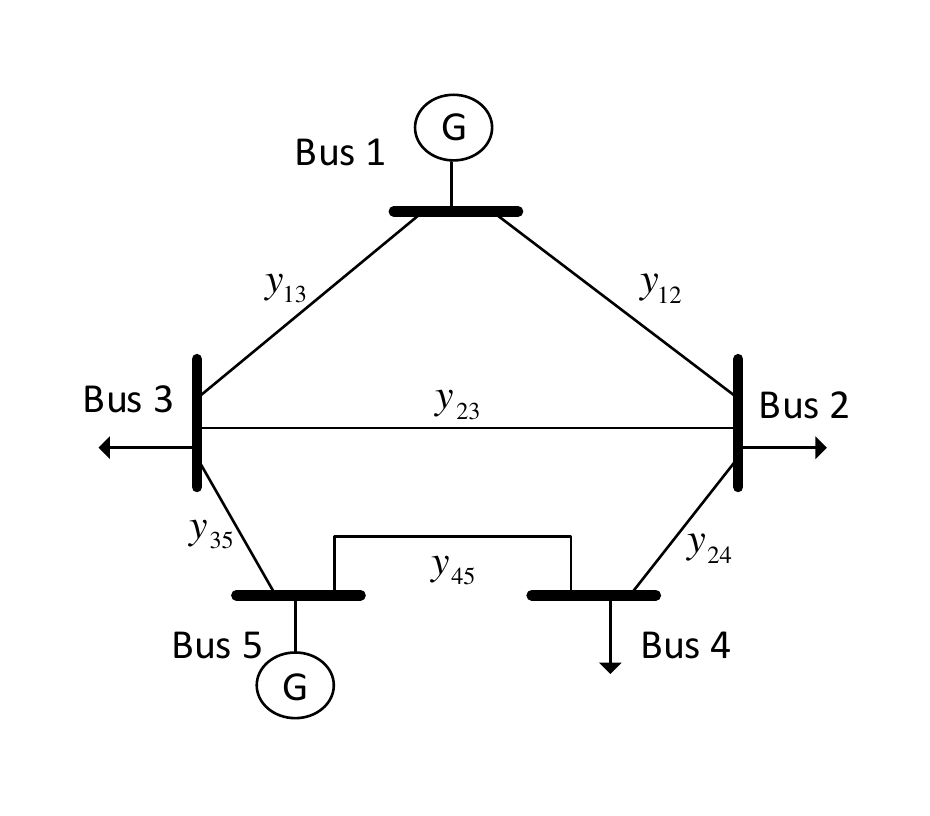}
\caption{WB5-Five Bus Network \cite{BGMT_webpage}}
\label{network}
\end{figure}

For illustrative purpose, Figure \ref{network} provides a diagram of
WB5-Five Bus Network \cite{BGMT_webpage} with $\clN=\{1, 2, 3, 4, 5\}$. It is seen from the bus connection
in this Figure that $\clN(1)=\{2,3\}$ because buses $2$ and $3$ are connected to bus $1$,
$\clN(2)=\{ 1, 3, 4\}$, $\clN(3)=\{1, 5\}$, $\clN(4)=\{2,5\}$ and $\clN(5)=\{3,4\}$. Also $\clG=\{ 1,5\}$
because buses $1$ and $5$ are connected to generators.

The objective of OPF is to minimize either the total cost of the generated power defined by
\cite{DT68}
\begin{equation}\label{cost}
  f(P_{G}) = \ds \sum_{k\in {\mathcal G}}(c_{k2}P_{G_k}^2+c_{k1}P_{G_k}+c_{k0}),
\end{equation}
with given $c_{k2}>0$, $c_{k1}$ and $c_{k0}$ and real active generated power $P_{G}$, or
the total transmission losses defined by
\begin{equation}\label{loss}
F_{loss}(P_G)= \sum_{k\in {\mathcal G}} (P_{G_k} - P_{L_k})- \sum_{ k\in\clN\setminus\clG} P_{L_k}.
\end{equation}
In this paper, we consider only the objective function (\ref{cost}). It is quite clear that our approach can be adjusted to
solve the objective (\ref{loss}) in a straightforward manner.\\
Write the objective (\ref{cost}) as the following function of the bus voltages $V$:
\begin{eqnarray}
 f(V) = \ds \sum_{k\in {\mathcal G}}[c_{k2}(P_{L_k} + \Re (\ds \sum_{m\in {\mathcal N}(k)}V_k V_m^*y_{km}^*))^2\nonumber \\
 +c_{k1}(P_{L_k} + \Re (\ds \sum_{m\in {\mathcal N}(k)}V_k V_m^*y_{km}^*)) + c_{k0}].\label{objective2}
\end{eqnarray}
Accordingly, the  OPF problem is formulated as follows
\begin{subequations}\label{opf1}
\begin{align}
\ds\min_{V\in \mathbb{C}^n} f(V)\quad \mbox{s.t.}\label{opf1a}\\
-P_{L_k} - jQ_{L_k} = \ds \sum_{m\in {\mathcal N}(k)}V_k V_m^*y_{km}^*,
 k\in\clN\setminus\clG, \label{opf1b}\\
P_{G_k}^{min}\leq P_{L_k} + \Re (\ds \sum_{m\in {\mathcal N}(k)}V_k V_m^*y_{km}^*)\leq P_{G_k}^{max}, k\in \clG \label{opf1c}\\
Q_{G_k}^{min}\leq Q_{L_k} + \Im (\ds \sum_{m\in {\mathcal N}(k)}V_k V_m^*y_{km}^*)\leq Q_{G_k}^{max}, k\in \clG \label{opf1d}\\
V_k^{min}\leq |V_k| \leq V_k^{max},  k\in\clN, \label{opf1e}\\
|S_{km}|=|V_k V_m^*y_{km}^*| \leq S_{km}^{max}, \forall (k,m)\in\clL \label{opf1f}\\
|V_k-V_m|\leq V_{km}^{max},  (k,m)\in\clL, \label{opf1g}\\
|\mbox{arg}(V_k)-\mbox{arg}(V_m)| \leq \theta_{km}^{max}, (k,m)\in\clL, \label{opf1h}
\end{align}
\end{subequations}
where (\ref{opf1b}) is the equation of the balance between the demand and supply power at bus $k\in\clN\setminus\clG$,
(\ref{opf1c})-(\ref{opf1d}) are the power generation bounds
with the lower bound $P_{G_k}^{min}$, $Q_{G_k}^{min}$ and upper bound
$P_{G_k}^{max}$, $Q_{G_k}^{max}$ of the real power  reactive power generations,
 (\ref{opf1e}) are the voltage amplitude bounds,
 (\ref{opf1f})-(\ref{opf1h}) are capacity limitations with the line currents between the connected buses  constrained by (\ref{opf1f}), while  (\ref{opf1g})-(\ref{opf1h}) guarantee the voltage balance in terms of their magnitude and phases
\cite{ZMT11}.\\
One can see that (\ref{opf1}) is a highly nonlinear optimization subject to nonlinear inequality and equality constraints and thus poses a real challenge for its computation. A common approach is to
use the slack variables
\begin{equation}\label{varchange}
W_{km}=V_kV_m^*, k=1,...,n; m=1,...,n
\end{equation}
and then  recast the problem (\ref{opf1}) in $V\in \mathbb{C}^n$ to the following problem in
$W=[W_{km}]_{k,m=1,...,n}\in\mathbb{C}^{n\times n}$:
\begin{subequations}\label{opf2}
\begin{align}
\ds\min_{W\in\mathbb{C}^{n\times n}}\ F(W) \quad\mbox{s.t.}\quad W\succeq 0, \label{opf2a}\\
-P_{L_k} - jQ_{L_k} = \ds \sum_{m\in {\mathcal N}(k)}W_{km}y_{km}^*,\  k\in\clN\setminus\clG, \label{opf2b}\\
P_{G_k}^{min}\leq P_{L_k} + \Re (\ds \sum_{m\in {\mathcal N}(k)}W_{km}y_{km}^*)\leq P_{G_k}^{max},  \label{opf2c}\\
k\in \clG,\nonumber\\
  Q_{G_k}^{min}\leq Q_{L_k} + \Im (\ds \sum_{m\in {\mathcal N}(k)}W_{km}y_{km}^*)\leq Q_{G_k}^{max}, \label{opf2d}\\
  k\in \clG,\nonumber\\
 (V_k^{min})^2 \leq W_{kk} \leq (V_k^{max})^2, k\in \clN, \label{opf2e}\\
 |W_{km}y_{km}^*| \leq S_{km}^{max}, (k,m)\in\clL, \label{opf2f}\\
  W_{kk}+W_{mm}-W_{km}-W_{mk} \leq (V_{km}^{max})^2, \label{opf2g}\\
  (k,m)\in\clL,  \nonumber\\
  \Im({W_{km}}) \leq \Re({W_{km}})\tan \theta_{km}^{max}, (k,m)\in\clL, \label{opf2h}\\
  \rank(W)=1, \label{opf2i}
\end{align}
\end{subequations}
where $F(W)=\ds\sum_{k\in {\mathcal G}}[c_{k2}(P_{L_k} + \Re (\ds \sum_{m\in {\mathcal N}(k)}W_{km}y_{km}^*))^2\\
+c_{k1}(P_{L_k} + \Re (\ds \sum_{m\in {\mathcal N}(k)}W_{km}y_{km}^*)) + c_{k0}]$,
which is convex quadratic in $W_{km}$, while all constraints (\ref{opf2b})-(\ref{opf2h}) are linear.
All the problem nonconvexity is now concentrated in the single rank-one constraint (\ref{opf2i}),
which is then dropped for SDR. If the solution of this SDR is of rank-one, then
it obviously leads to the global solution of the nonconvex optimization problem (\ref{opf2}). Otherwise even a feasible point
of (\ref{opf2}) is hardly retrieved from the SDR solution. Moreover formulation (\ref{opf2})
cannot be practically used for $n$ more than a few thousands in large-scale networks as it involves
{\color{black}$n(n+1)/2$ variables--a prohibitively large number.} The next section will address and resolve all these issues.
\section{Nonsmooth optimization based solution}
The first issue is to decompose large-size matrix $W$ in (\ref{opf2}) into matrices of smaller size
{\color{black} to make the problem tractable even with limited computational power.  }
This is also prompted by the
fact that there is only a small portion of the crossed terms $V_kV_m^*$
appearing in the nonlinear constraints (\ref{opf1b})-(\ref{opf1h})
so the large-size matrix variable $W\in\mathbb{C}^{n\times n}$ contains many redundant terms $V_kV_m^*$. The main result of
\cite{Metal13,AHV14,MAL15} is to decompose the set $\clN:=\{1, 2,...,n\}$ of buses
into $\clI$ overlapped subsets
$\clN_{i}=\{i_1,...,i_{N_i}\}$ of buses, {\color{black}called bags}, such that $
i_{\ell}\in \clN(i_{\ell+1}), \ell=1,...,N_i-1 \quad\mbox{and}\quad i_{N_i}\in\clN(i_1),
$ for each $i=1, 2,...,\clI$,
i.e. the buses in the same bag are serially connected. The set of bags can be reset to make bags of
relatively same size. Define the Hermitian symmetric matrix
variables
\begin{equation}\label{lvar}
W^{i}=[W_{i_ki_m}]_{k,m=1,..,N_i}\in\mathbb{C}^{N_i\times N_i}, i=1, 2,...,\clI.
\end{equation}
By replacing $W_{km}=V_kV_m^*$ in (\ref{opf2})  we have the
following equivalent  reformulation for (\ref{opf2})
\begin{subequations}\label{opfl}
\begin{eqnarray}
&\ds\min_{W={\sf diag}\{W^i\}}\ F(W)\quad \mbox{s.t.}\quad (\ref{opf2b})-(\ref{opf2h}), &\label{opfla}\\
& W^i\succeq 0,   i=1, ...,\clI,&\label{opf2i}\\
& {\sf rank}(W^i)=1, i=1, ...,\clI.&\label{opflb}
\end{eqnarray}
\end{subequations}
Reference \cite{Metal13,AHV14} just dropped all rank-one constraints in (\ref{opflb}) for SDR without any
justification. Reference \cite{MAL15} also dropped all rank-constraints in (\ref{opflb})
but then used a penalized SDR for locating low-rank semi-definite matrices $W^{i}$ in (\ref{lvar}). Based on these
low-rank matrices, \cite[Sec. IV]{MAL15} also proposed to find rank-one matrices, which however are not necessarily feasible to
(\ref{opfl}). \\
The variable number in (\ref{opfl}) is $\sum_{i=1}^{\clI}N_i(N_i+1)/2$. To keep this number reasonably moderate, it is desired that both
$\clI$ and $N_i$ are sufficiently moderate.   However, one can see that the above described decomposition \cite{Metal13,AHV14,MAL15} leads to a large number $\clI$ of bags as well as few large size $N_i$ that result in many rank-one constraints in (\ref{opflb}),
which are much less probably satisfied by solving SDR.

Our first step toward to computation  of (\ref{opf1}) is to develop a new decomposition with many fewer bags
involved. Recalling that $\clN(k)$ is the set of the buses that are connected to bus $k$, the cardinality
$|\clN(k)|$ is small in large-scale networks. We resort $\clN=\{1,2...,n\}$ as $\clN=\{N_1,...,\\N_n\}$
such that the cardinality $|\clN(N_k)|$ is in decreased order:
\[
|\clN(N_1)|\geq |\clN(N_2)|\geq ...\geq |\clN(N_n)|.
\]
Accordingly, the first bag of buses is defined as $\clN_1=\clN(N_1)$.
The second bag is defined as
\[
\clN_2=\{i\in \clN(N_2):\  \{i,N_2\} \not\subset \clN_1\}.
\]
Note that the crossed term $V_iV_{N_2}^*$ is already treated in the previous bag $\clN_1$
whenever $\{i,N_2\} \subset \clN_1$   so we exclude such bus $i$ in defining bag $\clN_2$. \\
Similarly, for $\ell\geq 3$ the $\ell$-th bag is defined as
\[
\clN_{\ell}=\{i\in\clN(N_{\ell}):  \{i,N_{\ell}\}\not\subset \clN_{\ell'}\ \forall 1\leq \ell'\leq \ell-1\}
\]
to exclude those buses $i$, whose crossed term $V_iV_{N_{\ell}}^*$ already is treated in a previous bag.\\
As each $|\clN_i|$ is obviously small,  such decomposition is
very efficient, leading to a substantial reduction of involved bags in comparison with that
used in \cite{Metal13,AHV14,MAL15}.

Our next step is to tackle the numerous difficult  rank-one constraints in
(\ref{opflb}), not dropping them for SDR as in all the previous works.\\
Firstly we express $\clI$ rank-one constraints in (\ref{opflb}) by the following single spectral constraint
\begin{equation}\label{spec}
\sum_{i=1}^{\clI}(\Tr(W^i)-\lambda_{\max}(W^i))=0,
\end{equation}
where $\lambda_{\max}(W^{i})$ stands for the maximal eigenvalue of $W^{(i)}$.
Indeed, (\ref{opf2i}) implies $\Tr(W^i)-\lambda_{\max}(W^i)\geq 0\ \forall\ i$, so (\ref{spec}) means that
$\Tr(W^i)=\lambda_{\max}(W^i)$, i.e. $W^{i}$ has only one nonzero eigenvalue so it is of rank-one.
The nonnegative quantity $\sum_{i=1}^{\clI}(\Tr(W^i)-\lambda_{\max}(W^i))$  can therefore be used to measure the degree of satisfaction of the rank-one constraints  (\ref{spec}). Without squaring, the penalization $\sum_{i=1}^{\clI}(\Tr(W^i)-\lambda_{\max}(W^i))$ is exact, meaning that (\ref{spec}) can be satisfied by a minimizer of
the problem
\begin{align}
\ds\min_{W={\sf diag}\{W^i\}}\ F_{\mu}(W):=F(W)+\mu\ds\sum_{i=1}^{\clI}(\Tr(W^i)\nonumber\\
-\lambda_{\max}(W^i))\quad \mbox{s.t.}\quad (\ref{opf2b})-(\ref{opf2h}), (\ref{opf2i}),\label{W3}
\end{align}
with a finite value of $\mu>0$ (see e.g. \cite[Chapter 16]{Betal06}). This is generally considered
as a sufficiently nice property to make such exact penalization attractive. \\
{\color{black}For any $W^{i,(\kappa)}$ feasible for the convex constraints (\ref{opf2b})-(\ref{opf2h}), (\ref{opf2i}),} function $\lambda_{\max}(W^i)$ is nonsmooth and is lower bounded by
\begin{equation}
\lambda_{\max}(W^i)=\ds\max_{||w||=1}w^HW^iw\geq (w_{\max}^{i,(\kappa)})^HW^iw_{\max}^{i,(\kappa)},
\label{W4}
\end{equation}
where $w^{i,(\kappa)}_{\max}$ is the normalized eigenvector corresponding to the eigenvalue $\lambda_{\max}(W^{i,(\kappa)})$,
i.e.
\begin{equation}
\lambda_{\max}(W^{i,(\kappa)})=(w_{\max}^{i,(\kappa)})^HW^{i,(\kappa)}w_{\max}^{i,(\kappa)}.
\end{equation}
{\color{black}Accordingly, $\mu\lambda_{\max}(W^i)-\mu\lambda_{\max}(W^{i,(\kappa)})\geq
\mu((w_{\max}^{i,(\kappa)})^HW^i\\ .w_{\max}^{i,(\kappa)}-(w_{\max}^{i,(\kappa)})^HW^{i,(\kappa)}w_{\max}^{i,(\kappa)})
=
\mu \la w_{\max}^{i,(\kappa)}(w_{\max}^{i,(\kappa)})^H, W^i-W^{i,(\kappa)}\ra$,
so $\mu w_{\max}^{i,(\kappa)}(w_{\max}^{i,(\kappa)})^H$ is a subgradient of the function
$\mu\lambda_{\max}(W^i)$ at $W^{i,(\kappa)}$. Then
$\mu{\sf diag}\{w_{\max}^{i,(\kappa)}(w_{\max}^{i,(\kappa)})^H\}$ is a subgradient of the function
$\mu\sum_{i=1}^{\clI}\lambda_{\max}(W^i)$ at ${\sf diag}\{W^{i,(\kappa)}\}$.
}\\

The following SDP  provides
an upper bound for {\color{black}the}  nonconvex {\color{black}optimization problem} (\ref{W3})
\begin{align}
\ds\min_{W={\sf diag}\{W^i\}}\ F^{(\kappa)}(W):=F(W)+\mu\ds\sum_{i=1}^{\clI}(\Tr(W^i)\nonumber\\
-(w_{\max}^{i,(\kappa)})^HW^iw_{\max}^{i,(\kappa)}) \quad\mbox{s.t.}\quad (\ref{opf2b})-(\ref{opf2h}), (\ref{opf2i})
\label{W5}
\end{align}
because $F^{(\kappa)}({\sf diag}\{W^i\})\geq F_{\mu}({\sf diag}\{W^i\})\ \quad\forall\ W^i\succeq 0$
according to (\ref{W4}). Suppose that $W^{(\kappa+1)}={\sf diag}\{W^{i, (\kappa+1)}\}$ is the optimal solution of SDP (\ref{W5}). Since $W^{(\kappa)}={\sf diag}\{W^{i,(\kappa)} \}$ is also
feasible to (\ref{W5}) with $F_{\mu}(W^{(\kappa)})=F^{(\kappa)}(W^{(\kappa)})$, it is true that
\[F_{\mu}(W^{(\kappa+1)})\leq  F^{(\kappa)}(W^{(\kappa+1)})\leq F^{(\kappa)}(W^{(\kappa)})
= F_{\mu}(W^{(\kappa)}),
\]
so $W^{(\kappa+1)}$ is a better feasible point of (\ref{W3}) than $W^{(\kappa)}$.
Initialized by any feasible point $W^{(0)}={\sf diag}\{W^{i,(0)}\}$ of SDP constraint
 (\ref{opfla})-(\ref{opf2i}),
the sequence $\{W^{(\kappa)}\}=\{ {\sf diag}\{W^{i,(\kappa)}\}\}$ with $W^{(\kappa+1)}={\sf diag}\{W^{i,(\kappa+1)}\}$ iteratively generated as the optimal solution of SDP (\ref{W5})
is a sequence of improved feasible points of the nonconvex optimization problem (\ref{W3}).
Since $W^{(\kappa)}$ are uniformly bounded, the sequence $\{W^{(\kappa)}\}$ has
a limit point
$\bar{W}={\sf diag}\{\bar{W}^i\}$, which  is the optimal solution of the optimization problem
\begin{align}
\ds\min_{W={\sf diag}\{W^i\}}\ F(W)+\mu\ds\sum_{i=1}^{\clI}(\Tr(W^i)\nonumber\\
-(\bar{w}_{\max}^{i})^HW^i\bar{w}_{\max}^{i}) \quad\mbox{s.t.}\quad (\ref{opf2b})-(\ref{opf2h}), (\ref{opf2i}),
\label{pW5}
\end{align}
where $\bar{w}^{i}_{\max}$ is the normalized eigenvector corresponding to the eigenvalue $\lambda_{\max}(\bar{W}^{i})$
of $\bar{W}^i$. Particularly,
\[
\begin{array}{ll}
F(W)+\mu\ds\sum_{i=1}^{\clI}(\Tr(W^i)-(\bar{w}_{\max}^{i})^HW^i\bar{w}_{\max}^{i})&\geq\\
F(W)+\mu\ds\sum_{i=1}^{\clI}(\Tr(\bar{W}^i)-(\bar{w}_{\max}^{i})^H\bar{W}^i\bar{w}_{\max}^{i}),
\end{array}
\]
or equivalently, under the definition
$g(W)=F(W)+\mu\ds\sum_{i=1}^{\clI}\Tr(W^i)$,
\[
g(W)-g(\bar{W})-
\la \mu{\sf diag}\{\bar{w}_{\max}^i(\bar{w}_{\max}^i)^H\},
W-\bar{W}\ra \geq 0
\]
for all feasible points $W={\sf diag}\{W^i\}$ in (\ref{opf2b})-(\ref{opf2h}), (\ref{opf2i}).
As a result, $\bar{W}$ is the optimal solution of the convex optimization problem
\[
\begin{array}{r}
\ds\min_{W={\sf diag}\{W^i\}}\ g(W)-\la \mu{\sf diag}\{\bar{w}_{\max}^i(\bar{w}_{\max}^i)^H \},
W-\bar{W}\ra\\
\mbox{s.t.}\quad (\ref{opf2b})-(\ref{opf2h}), (\ref{opf2i}),
\end{array}
\]
so it must satisfy the optimality condition
\[
\la \nabla g(\bar{W})-\mu{\sf diag}\{\bar{w}_{\max}^i(\bar{w}_{\max}^i)^H \},
W-\bar{W}\ra\geq 0
\]
for all feasible points ${\sf diag}\{W^i\}$ in (\ref{opf2b})-(\ref{opf2h}), (\ref{opf2i}). The latter is also the
first order necessary optimality condition for {\color{blue}(\ref{W3})} because $\mu{\sf diag}\{\bar{w}_{\max}^i(\bar{w}_{\max}^i)^H \}$
is a subgradient of the function $\mu\sum_{i=1}^{\clI}\lambda_{\max}(W^i)$ at $\bar{W}$.
As our simulations will show, $\bar{W}$ is indeed the global optimal solution of (\ref{W3}) and
(\ref{opfl}).

However, unlike \cite{STST15} with only a single rank-one constrained
matrix, although quantity
\begin{equation}\label{sum}
\sum_{i=1}^{\clI}(\Tr(W^{i, (\kappa)})-\lambda_{\max}(W^{i, (\kappa)}))
\end{equation}
in (\ref{W3}) is iteratively decreased, not all individual quantities
\begin{equation}\label{term}
\Tr(W^{i, (\kappa)})-\lambda_{\max}(W^{i, (\kappa)})
\end{equation}
are iteratively decreased so
the rank of each matrix $W^{i,(\kappa)}$ is no longer iteratively reduced to one as expected. Worse,
$W^{i,(\kappa)}$ is rank-one  but the rank of  $W^{i,(\kappa+1)}$ in the next iteration may turn to be more than one with
\[
\begin{array}{ll}
\Tr(W^{i, (\kappa+1)})-\lambda_{\max}(W^{i, (\kappa+1)})&>\\
\Tr(W^{i, (\kappa)})-\lambda_{\max}(W^{i, (\kappa)}).\end{array}
\]
Consequently, it is very difficult to achieve rank-one for all  $W^{i,(\kappa)}$
as desired. It is also impossible to  add
a "weight" to each term under the sum in the objective
in (\ref{W5}) to handle the individual convergence of $\Tr(W^{i})-\lambda_{\max}(W^{i})$.\\
We now develop a systematic way to resolve this issue as follows. For $\kappa=0, 1,...,$ and $W^{(\kappa)}={\sf diag}\{W^{(i,(\kappa))}\}$ define
\begin{equation}\label{goodk}
\clL^{(\kappa)}=\{i\in\{1,...,\clI\} \ :\  {\sf rank}(W^{i,(\kappa)})=1\}
\end{equation}
and generate $W^{(\kappa+1)}={\sf diag}\{W^{i,(\kappa+1)}\}$ as the optimal solution of the
following SDP instead of SDP (\ref{W5})
\begin{subequations}\label{Wl5m1}
\begin{eqnarray}
\ds\min_{W={\sf diag}\{W^i\}}\ F(W)+\mu{\color{black}\sum_{i=1}^{\clI}}[\Tr(W^i)\nonumber\\
-(w_{\max}^{i,(\kappa)})^HW^iw_{\max}^{i,(\kappa)}]
\quad\mbox{s.t.}\quad (\ref{opf2b})-(\ref{opf2h}), (\ref{opf2i}),\label{W15m1a}\\
\Tr(W^i)-(w_{\max}^{i,(\kappa)})^HW^iw_{\max}^{i,(\kappa)}\leq \epsilon_{tol},
i\in\clL^{(\kappa)}.\label{W15m1b}
\end{eqnarray}
\end{subequations}
Note that $\Tr(W^{i})\geq w^HWw$ for all  $||w||=1$ and it is obvious that
$\rank(W^{i})=1$ if and only if $\Tr(W^{i})-w_{max}^HW^{i}w_{\max}=0$ for some normalized $w_{\max}$. Therefore,
the constraint (\ref{W15m1b}) for some tolerance $\epsilon_{tol}$ is introduced to warrant the rank-one of
all $W^{i,(\kappa+1)}$, $i\in\clL^{(\kappa)}$. As a result $\clL^{(\kappa)}\subset \clL^{(\kappa+1)}$ and
$\clL^{(\kappa)}\rightarrow  \{1,...,\clI\}$ is expected to have all $W^{i,(\kappa)}$ of rank-one.
Unlike (\ref{W5}), the iterations (\ref{Wl5m1}) leads to achieving rank-one of all $W^i$
while the objective function $F_{\mu}$ is still decreased.

In summary, we propose the following {\it Large-Scale Non-smooth Optimization Algorithm (Large-scale NOA) }
 for the multiple rank-one constrained optimization problem (\ref{W3}).\\

{\it Initialization.} Solve SDP
\begin{equation}\label{W13}
\ds\min_{W={\sf diag}\{W^i\}}\ F(W) \quad\mbox{s.t.}
\quad (\ref{opf2b})-(\ref{opf2h}), (\ref{opf2i})
\end{equation} to generate
$W^{(0)}:={\sf diag}\{W^{i,(0)}\}$. If
$\rank(W^{i,(0)})\equiv 1$ stop: $W^{(0)}$ is the global solution of
the nonconvex optimization problem
(\ref{opfl}).  Otherwise set $\kappa=0$ and define $\clL^{(\kappa)}$ by (\ref{goodk}).

{\it $\kappa$-th iteration.} For $\kappa=0, 1,..,$ solve (\ref{Wl5m1})  to generate $W^{(\kappa+1)}:={\sf
diag}\{W^{i,(\kappa+1)}\}$. Reset $\kappa=\kappa+1$ and define $\clL^{(\kappa)}$ by (\ref{goodk}).
Stop whenever $\clL^{(\kappa)}=\{1,...,\clI\}$. Otherwise go to the next iteration.
\vspace*{-0.2cm}
\section{Simulation results}
\vspace*{-0.2cm}
The computation facilities for our implementation are
Processor Intel(R) Core i5-3470 CPU @3.20GHz, Matlab version R2013b and
CVX with SDPT3. We set the tolerances $\epsilon=\epsilon_{tol} = 10^{-5}$ and
the penalty parameter $\mu=10^{6}$ which
makes the penalty term $\mu\sum_{i=1}^{\cal I}(\Tr(W^{(i,(0))})-\lambda_{\max}(W^{i, (0)}))$   at similar magnitude
with the objective $F(W^{(0)})$.
The data source for all examples is  Matpower version 5.1 \cite{ZMT11}. {\color{black}All examples were considered
in \cite{Metal13,AHV14,MAL15} by SDR only.} We recall that $\clT$ is the number of matrix variables $W^{i}$ in the OPF problem
(\ref{opfl}) and $\clL^{(\kappa)}$ is defined by (\ref{goodk}) is the set of
rank-one matrices $W^i$ found after $\kappa$-th iteration.  The capability of our large-scale NOA in locating
the global optimal solution of the OPF problems is demonstrated by showing that the global optimality tolerance (GOT)
of its found solution defined as
\[
\frac{\mbox{\sf the found value- lower bound}}
{\mbox{\sf lower bound}}
\]
is almost zero.

The numerical examples are presented as follows.
\subsection{Polish-2383wp system}
There are $n=2383$ buses, $327$ generators and $2896$ transmission lines, leading to $2056$ nonlinear constraints in (\ref{opf1b}).\\
{\it Initialization.} A lower bound $1.8490\times 10^6$ of (\ref{opfl}) is found by solving SDP  (\ref{W13}).
$|\clL^{(0)}|=1210$ and there are $32$ matrices $W^{i,(0)}$ of rank-more-than-one. Their largest
size (smallest size, resp.) is $10\times 10$ ($2\times 2$, resp.).\\
{\it Stage 1.}  {\color{black}$|\clL^{(10)}|=1234$} is achieved. There are {\color{black}$8$} matrices
{\color{black}$W^{i,(10)}$} of rank-more-than-one.
  Their largest size (smallest size, resp.) is $10\times 10$ and
  ($3\times 3$, resp.).\\
{\it Stage 2.}  {\color{black}$|\clL^{(19)}|=1237$} is achieved. There are $5$ matrices $W^{i,(20)}$ of rank-more-than-one.
Their largest size (smallest size, resp.) is $9\times 9$
($3\times 3$, resp.).\\
{\it Stage 3.} {\color{black}$|\clL^{(25)}|=\clI=1242$} is achieved.
The found  value of the  objective is {\color{black}$1.8408\times 10^6$} with GOT {\color{black}$4.3267e-04$.}

\begin{table*}[h]
\centering
\footnotesize
\caption{Comparison of bags number $\clI$, largest bag size $M_i$
and number of variables}
\resizebox{\textwidth}{!}{%
\label{t1}
\scalebox{0.7}{
\begin{tabular}{|c|c|c|c|c|c|c|c|c|c|}
\hline
System  & $\clI$ & $\clI$ by \cite{MAL15} & Max. $N_i$ & Max. $N_i$ by \cite{MAL15} &
Var. \#  in (\ref{opf2}) &Var. \#  in (\ref{opfl}) &Var. \#  in (\ref{opfl}) by \cite{MAL15} & Found value & Found by \cite{MAL15}\\ \hline
Polish-2383wp  & 1242        & 2383                   & 10               & 23                          & 2,840,536                & 23,199                  & 89,893                             & ${\color{black}1.8408}\times 10^6$ & $1.8742\times 10^6$ \\ \hline
Polish-2736sp  & 1538        & 2736                   & 10               & 23                          & 3,744,216                & 27,298                  & 104,388                            & $1.3042\times 10^6$ & $1.3082\times 10^6$ \\ \hline
Polish-2737sop & 1538        & 2737                   & 10               & 23                          & 3,746,953                & 27,034                  & 103,720                            & $7.7572\times 10^5$ & $7.7766\times 10^5$ \\ \hline
Polish-2746wop & 1546        & 2746                   & 10               & 23                          & 3,771,631                & 29,024                  & 108,950                            & $1.2040\times 10^6$ & $1.2085\times 10^6$ \\ \hline
Polish-2746wp  & 1547        & 2746                   & 10               & 24                          & 3,771,631                & 28,257                  & 107,148                            & $1.6266\times 10^6$ & $1.6324\times 10^6$ \\ \hline
Polish-3012wp  & 1689        & 3012                   & 10               & 24                          & 4,537,578                & 30,996                  & 116,799                            & $2.5727\times 10^6$ & $2.6089\times 10^6$ \\ \hline
Polish-3120sp  & 1757        & 3120                   & 10               & 24                          & 4,868,760                & 32,637                  & 121,869                            & $2.1391\times 10^6$ & $2.1608\times 10^6$ \\ \hline
\end{tabular}}}%
\end{table*}

\begin{table*}[h]
\centering
\footnotesize
\caption{Performance comparison}
\label{t2}
\begin{tabular}{|c|c|c|c|}
\hline
System & Found value & Found by \cite{MAL15}&{\color{black}Found by \cite{ZMT11}}\\ \hline
Polish-2383wp   & ${\color{black}1.8408}\times 10^6$ & $1.8742\times 10^6$&{\color{black}$1.8685\times 10^6$} \\ \hline
Polish-2736sp  & $1.3042\times 10^6$ & $1.3082\times 10^6$&{\color{black} $1.3078\times 10^6$} \\ \hline
Polish-2737sop & $7.7572\times 10^5$ & $7.7766\times 10^5$&{\color{black} $7.7763\times 10^5$} \\ \hline
Polish-2746wop &  $1.2040\times 10^6$ & $1.2085\times 10^6$&{\color{black} $1.2083\times 10^5$} \\ \hline
Polish-2746wp  &  $1.6266\times 10^6$ & $1.6324\times 10^6$&{\color{black} $1.6317\times 10^6$} \\ \hline
Polish-3012wp  &  $2.5727\times 10^6$ & $2.6089\times 10^6$ &{\color{black} $2.5917\times 10^6$}\\ \hline
Polish-3120sp  & $2.1391\times 10^6$ & $2.1608\times 10^6$&{\color{black} $2.1427\times 10^6$} \\ \hline
\end{tabular}%
\end{table*}
\subsection{Polish-2736sp system}
There are $n=2736$ buses, $420$ generators and $3504$ transmission lines, which lead to $2316$ nonlinear constraints in (\ref{opf1b}). \\
{\it Initialization.} A lower bound $1.3041\times 10^6$ of (\ref{opfl}) is obtained by solving
SDP (\ref{W13}). $|\clL^{(0)}|=1534$ and there are $4$ matrices $W^{i,(0)}$ of rank-more-than-one.  Their largest size (smallest size, resp.)  is $6\times 6$ ($4\times 4$, resp.).\\
{\it Stage 1.} $|\clL^{(9)}|=\clI=1538$ is achieved.
The found  value of the  objective is $1.3042\times 10^6$ with GOT $7.6681e-05$.
\subsection{Polish-2737sop system}
There are $n=2737$ buses, $399$ generators and $3506$ transmission lines, which lead to $2338$ nonlinear constraints in (\ref{opf1b}). \\
{\it Initialization.} A lower bound $7.7571\times 10^5$ of  (\ref{opfl}) is obtained by solving
SDP (\ref{W13}). $|\clL^{(0)}|=1532$ and there are $6$  matrices $W^{i,(0)}$ of rank-more-than-one.  Their largest size (smallest size, resp.) is $6\times 6$ ($3\times 3$, resp.).\\
{\it Stage 1.}  {\color{black}$|\clL^{(2)}|=\clI=1538$} is achieved.
The found  value of the  objective is $7.7572\times 10^5$ with GOT $1.2891e-05$.
\subsection{Polish-2746wop system}
There are $n=2746$ buses, $514$ generators and $3514$ transmission lines, which lead to $2232$ nonlinear constraints in (\ref{opf1b}). \\
{\it Initialization.} A lower bound $1.2039\times 10^6$
of  (\ref{opfl}) is obtained by solving SDP (\ref{W13}).
$|\clL^{(0)}|=1538$  and there are $8$ matrices $W^{i,(0)}$ of rank-more-than-one.  Their largest size (smallest size, resp.) is $6\times 6$ ($3\times 3$, resp.).\\
{\it Stage 1.} $|\clL^{(2)}|=\clI=1546$ is achieved.
The found  value of the  objective is $1.2040\times 10^6$ with GOT $8.3063e-05$.
\subsection{Polish-2746wp system}
There are $n=2746$ buses, 520 generators and 3514 transmission lines, which lead to 2226 nonlinear constraints in (\ref{opf1b}).\\
{\it Initialization.} A lower bound  $1.626590\times 10^6$ of  (\ref{opfl})
is obtained by solving SDP (\ref{W13}).
$|\clL^{(0)}|=1545$  and there are $2$ matrices $W^{i,(0)}$ of rank-more-than-one.  Their size is $4\times 4$.\\
{\it Stage 1.} $|\clL^{(1)}|=\clI=1547$ is achieved.
The found  value of the  objective is {\color{black}$1.626591\times 10^6$} with GOT $6.1478e-07$.
\subsection{Polish-3012wp system}
There are $n=3012$ buses, 502 generators and 3572 transmission lines, which lead to 2510 nonlinear constraints in (\ref{opf1b}). \\
{\it Initialization.} A lower bound $2.5717\times 10^6$ of (\ref{opfl})
is obtained by solving SDP  (\ref{W13}).
$|\clL^{(0)}|=1682$ and there are $7$ matrices $W^{i,(0)}$ of rank-more-than-one.
Their largest size (smallest size, resp.) is $7\times 7$ ($2\times 2$, resp.).\\
{\it Stage 1.} {\color{black}$|\clL^{(4)}|=\clI=1689$} is achieved.
The found  value of the  objective is $2.5727\times 10^6$ with GOT $3.8885e-04$.
\subsection{Polish-3120sp system}
There are $n=3120$ buses, 505 generators and 3693 transmission lines, which lead to 2615 nonlinear constraints in (\ref{opf1b}). \\
{\it Initialization.} A lower bound $2.1314\times 10^6$ of (\ref{opfl}) is
obtained by solving SDP (\ref{W13}).
$|\clL^{(0)}|=1749$  and  there are $8$  matrices $W^{i,(0)}$ of rank-more-than-one.
Their largest size (smallest size, resp.) is $8\times 8$ ($2\times 2$, resp.).\\
{\it Stage 1.} $|\clL^{(9)}|=\clI=1757$ is achieved.
The found  value of the  objective is $2.1391\times 10^6$ with GOT $0.0036$.
\subsection{Numerical summary}
One can observe that GOT of the solutions computed by the large-scale NOA is very small,
proving its capability  to provide the global solution of (\ref{opfl}).
Table \ref{t1} and Table \ref{t2} summarize the main points in our simulation.
The second and third columns of Table \ref{t1}
are the number $\clI$ of bags  in (\ref{lvar}) by our decomposition and by that in \cite{MAL15}, while
the fourth and fifth columns give the maximum size $N_i$ in (\ref{lvar}). One can see that both $\clI$ and the maximum
$N_i$ by our decomposition are substantially smaller than their counterparts by \cite{MAL15}.
This leads to far smaller numbers of variables in
(\ref{opfl}), which are provided in the seventh and eighth columns.
The number $n(n+1)/2$ of complex variables in (\ref{opf2}) is also provided in the
sixth column to contrast to the number of complex variables in (\ref{opfl}) in the seventh column.
{\color{black}Furthermore, the second column of Table \ref{t2} provides
the best values of (\ref{opf1}) found by our large-scale NOA, which are far smaller than ones in the third
and fourth columns found  by \cite{MAL15} and Matpower6.0 \cite{ZMT11} (using an interior point method), respectively. }
In short, our computation approach to the OPF problem (\ref{opfl}) outperforms  other existing
approaches in terms of computational efficiency and performance.
\vspace*{-0.2cm}
\section{Conclusion}
The OPF problems over power transmission network are large-scale  optimization problems, which involve a large number of
quadratic equality and indefinite quadratic inequality constraints and thus are difficult computationally.
We have developed a large-scale nonsmooth optimization algorithm  to
compute their  optimal solutions, which is efficient and practical for large-scale power transmission networks of a few thousands of buses. Applications of the developed large-scale NOA to the OPF problems over three-phase power transmission networks are currently under investigation.\\

{\bf Acknowledgements.} We thank Cedric Josz, a coauthor of \cite{J1} and \cite{J2} for sending us these references.
\bibliography{OPF_BIB}%
\end{document}